# Prospective Validation of Motor-Based Intervention with Automated Mispronunciation Detection of Rhotics in Residual Speech Sound Disorders

*Nina R Benway[1], Jonathan L Preston[1,2]*

[1]Communication Sciences and Disorders, Syracuse University, New York, USA
[2]Haskins Laboratories, New Haven, Connecticut, USA
`nrbenway@syr.edu`

## Abstract

Because lab accuracy of clinical speech technology systems may be overoptimistic, clinical validation is vital to demonstrate system reproducibility–in this case, the ability of the PERCEPT-R Classifier to predict clinician judgment of American English /ɹ/ during ChainingAI motor-based speech sound disorder intervention. All five participants experienced statistically-significant improvement in untreated words following 10 sessions of combined human-ChainingAI treatment. These gains, despite a wide range of PERCEPT-human and human-human (F1-score) agreement, raise questions about best measuring classification performance for clinical speech that may be perceptually ambiguous.

## 1. Introduction

Technologies automating clinically-validated evidence-based practices for speech sound intervention are well suited to tackle worldwide barriers to sufficiently intense intervention [1]. We focus on mispronunciation detection for /ɹ/, the most common American English speech sound disorder. Our recent work has offset several factors [2-5] previously precluding the development of effective clinical speech technologies. First, the PERCEPT-R Corpus [6] confronts the *paucity of (labeled) public child/clinical speech corpora*. Second, *adequate technical documentation* of the PERCEPT-R Classifier [7], a gated recurrent neural network, reports participant-specific F1-score $\bar{x}$ = .81 ($\sigma_x$ = .10; med = .83, n = 48) for the detection of fully rhotic versus derhotic /ɹ/ in speech sound disorder. Lastly, the present study reports *prospective clinical validation* for intervention with the PERCEPT-R Classifier.

## 2. Related Work and Contributions

### 2.1. Speech Motor Chaining and ChainingAI

Speech Motor Chaining is an evidence-based, full-stack web app for speech sound intervention [8, 9] that is well-suited to deliver sufficiently intense intervention due to high fidelity of treatment replications and high sum of practice trials per session. Speech Motor Chaining modulates several *principles of motor learning* [10] to adapt practice difficulty based on a learner's performance and the short-term goal (i.e., motor *acquisition* or motor *learning*). Previous versions of Speech Motor Chaining require the clinician to key their judgment of a learner's practice attempt (i.e., fully rhotic/"correct" versus derhotic/ "incorrect") before the program calculates linguistic difficulty of the next trial. In contrast, *ChainingAI* predicts perceptual judgment of /ɹ/ with the PERCEPT Engine, which extracts age-and-sex normalized formants features from target intervals, identified by Gaussian mixture model/Hidden Markov models (GMM-HMM) [11], as described in [7]. These features are used by the PERCEPT-R Classifier to predict clinician perceptual judgment of the /ɹ/ and to drive Chaining practice adaptations. With the PERCEPT Engine, ChainingAI might help narrow the intervention intensity gap through, e.g., clinical-grade home practice between clinician-led sessions.

### 2.2. Prospective Validation Framework and Contributions

Prospective validation is required for reproducibility, as lab accuracy of clinical speech technologies may be overoptimistic [12]. Because the PERCEPT project meets several reproducibility guidelines [12] (low-dimension, validated acoustic phonetic features actively collected in relevant speech tasks), it is possible that PERCEPT's lab accuracy [7] may generalize to the clinic. However, this remains to be empirically demonstrated and necessitates careful prospective hardware verification, analytical validation, and clinical validation [13].

#### 2.2.1. Hardware Validation: Audio and Rhotic Capture

Audio capture in ChainingAI is controlled by start/stop buttons on the webpage in the learner's web browser, using the WebRTC getUserMedia() method. The entire target utterance must be captured by the user for successful GMM-HMM forced alignment. Even if captured, GMM-HMM acoustic models may lack validity for treatment speech; treatment-elicited /ɹ/ may not have a canonical steady state during establishment of a new motor plan, and is often elongated as a clinical strategy. Both factors may result in different distributions of HMM transition probabilities between lab speech (likely from word list reading) and speech (likely from treatment practice attempts). Our first contribution in this work shows that > 92% of audio is captured by the system, but GMM-HMM underestimates duration for word-initial rhotics in this sample of treatment-elicited speech.

#### 2.2.2. Analytical Validation: PERCEPT-R/Rater Reliability

It is clinically intuitive that automated clinical feedback should accurately reflect clinician perception; however, determining this ground truth is perhaps not so forthright. The perception of speech sounds, broadly, as mispronounced or not is not only informed by the phone itself but expectation of a sound's features and phonetic/phonotactic/lexical/prosodic contexts [14]. Indeed, previous lab ratings of /ɹ/ from children with speech sound disorders show 85% agreement between *expert* listeners [15]. Complicating reliability further at present is that the children for whom independent practice is clinically indicated—those with emerging /ɹ/—are those theorized to elicit the least reliable ground truths due to production of ambiguous/intermediate tokens compared to the more robust derhotic-rhotic endpoints seen during classifier training. Indeed, recent ratings of /ɹ/ in speech sound disorder yielded low inter-human reliability for ambiguous productions (ICC = .39, 95% CI [.22-.48]) [16], a conceptually-analogous feature space to the productions we expect herein. Indeed, our second contribution herein demonstrates a lower PERCEPT-human mode F1-score than lab testing, but individual PERCEPT-human F1-scores for all but one participant (averaging .61, $\sigma_x$= .14, n = 5 speakers) occupy a similar space compared to human-human F1-scores. This contribution raises interesting questions about ground truth generation for clinical speech.

*2.2.3. Clinical Validation: Participant Response to Treatment*

Previous studies of (clinician-led) Speech Motor Chaining show a large effect for improved /ɹ/ in untrained words after seven hours of treatment [8]. We hypothesize that ChainingAI can similarly foster improvements in /ɹ/ on untrained words, compared to no-treatment baselines. Because learner responses to state of the art speech therapies vary [17], with 5 participants in the experiment we anticipate that 3-4 participants will show change from pre-treatment to post-treatment that is significantly different than 0. Indeed, our third contribution shows a combined treatment program including ChainingAI facilitates speech motor learning for each of five study participants.

## 3. Methods

### 3.1. Experimental Design

This is a multiple baseline, single case experimental Phase II clinical trial [18] to prospectively validate the PERCEPT-R Classifier for in-treatment delivery of motor learning feedback. Ethical approval was granted by the institutional review boards of Syracuse University and The College of Saint Rose. All study visits, except "Orientation to /ɹ/", were video-conferenced and audio was recorded local to the participant. Note that experimental/treatment/user acceptability details of interest to clinical trialists are extensively detailed elsewhere [19].

### 3.2. Participants and PERCEPT-R Personalization

This report describes five participants (4 male, 1 female) with speech sound disorder aged 10-19 years ($\bar{x}$ = 12.7, $\sigma_x$ = 3.6). The imbalance between sexes is expected in Phase II trials, given the higher speech disorder prevalence in males [20]. The small sample is appropriate for a Phase II longitudinal trial and is a common sample size in single-case experiments for speech disorders [21]. The target enrollment was 5 and these five participants reflect the first meeting our a priori inclusionary criteria. The most important inclusionary characteristic was the ability to produce /ɹ/ some of the time (i.e., clinical *stimulability*), as we contend that learners must produce a correct /ɹ/ some of the time to benefit from practice without a clinician. Stimulability was defined as producing fully-rhotic /ɹ/ in > 20% of tested syllables but < 40% of words at study onset.

Participants completed a no-treatment baseline of randomized length (5-10 probes at a rate of 3 per week; and thrice post-treatment) in which word list recordings were made. In addition to evidencing the no-treatment comparison, baseline recordings were labeled (derhotic/fully rhotic) and age-and-sex normalized formant-based features were generated [7]. These features were used for participant-specific personalization of PERCEPT-R. For participants 1107, 1111, and 1112, the number of personalized layers in PERCEPT-R was set heuristically as the last linear layer and the output layer, with hyperparameters exactly as in [7]. For participants 1121 and 1130, the personalization procedure was updated such that the number of layers with freely-varying gradients (plus additional hyperparameters such as learning rate) were optimized through a warm-start search facilitated by Optuna. Personalization performance is summarized in Table 1. F1-score is the selected performance metric as it emphasizes clinically-desirable recall.

Table 1: *PERCEPT Baseline Performance (F1-score; [true derhotic, false rhotic | false derhotic, true rhotic], normalized by ground-truth label)*

| Participant | Out of Box | Personalized |
| --- | --- | --- |
| 1107 | .708 [.72, .28 | .30, .70] | .792 [.82, .18 | .24, .76] |
| 1111 | .383 [.19, .81 | 0, 1] | .780 [.73, .27 | .14, .86] |
| 1112 | .520 [.24, .76 | .12, .88] | .735 [.71, .29 | .24, .76] |
| 1121 | .614 [.83, .17 | .61, .39] | .808 [.69, .31 | .08, .92] |
| 1130 | .458 [.03, .97 | 0, 1] | .842 [.71, .29 | .05, .95] |

### 3.3. Treatment

Each participant had 10 motor-based treatment sessions, each 40 minutes, thrice/week. The first session was human-led "Orientation to /ɹ/", teaching articulatory gestures for fully-rhotic /ɹ/ in the context of motor learning prepractice [22]. This session ended with human-led overview of the web app interface. The next 9 sessions each had the same format: the first ≤ 10 minutes of each session were human-led prepractice, where syllable-level rhotic targets (i.e., "roo", "rah") were practiced with unlimited clinical instruction until tallying 4 fully rhotic productions per target. The remaining 30 minutes of each session was independent practice with ChainingAI, in which feedback and practice modulations were automated through the PERCEPT-R analysis of participant-recorded audio. Clinical feedback was delivered by the ChainingAI cartoon clinician. Human clinician involvement was limited to technical troubleshooting. The linguistic stages and difficulty adaptation algorithm of ChainingAI are described in [9, 19].

### 3.4. Outcome Measures

*3.4.1. Hardware Validation: Audio and Rhotic Capture*

Human listeners reviewed every study-collected utterance to determine the percent of files containing the complete target using the user-driven audio capture method and study-provided Shure Mv5 cardioid microphone. Next, the rhotic timestamps predicted by GMM-HMM (default English acoustic models) at runtime were compared to hand-corrected rhotic boundaries for each ChainingAI audio file. We examined the effect of GMM-HMM estimation by observing rhotic durations and the percent of ground-truth rhotic interval samples (1 ms steps) captured within the GMM-HMM interval.

*3.4.2. Analytical Validation: PERCEPT/Rater Reliability*

Each recording captured by ChainingAI was judged by expert raters having licensure in speech-language pathology and expertise in /ɹ/ speech sound disorders. Additionally, all raters completed training modules and passed a category goodness task [23] with ≤ 90% absolute agreement for classification of derhotic/fully rhotic tokens relative to a separate panel of experts. Each ChainingAI recording received a binary rhoticity rating from 3 of 7 panel members relative to a "treatment" standard (i.e., "would you have told the participant this was correct in a treatment session?"). This binary rating was derived from a 100-point visual analogue scale with anchors for *fully derhotic* (0), *fully rhotic* (100), and *clinically correct* (75). All raters were masked to participant ID and timepoint of utterance collection at the time of rating, with the order of utterance presentation randomized. We examined intra- and inter-human reliability of ratings for random sample of 5% of tokens per rater. We chose Gwet's chance-corrected agreement coefficient, which is more robust than kappa in the cases of high or low binary class prevalence [24]. We also report intraclass correlation coefficient (ICC) for comparison to previous literature. ICC was computed with the continuous visual analogue scale rating from which the binary rating was derived. Lastly, we also report F1-score compared to the mode of listener binary ratings for each individual as well as pairwise comparison of F1-score for the human raters and PERCEPT. For human-human comparisons, the F1-score is reported as an average of two calculations, allowing both humans to serve as ground truth. For human-PERCEPT comparisons, only human perceptual judgment served as ground truth for F1-score.

### 3.4.3. Clinical Validation: Response to Treatment

To assess clinical change, recordings of untreated rhotic words were collected in the week immediately following the completion of treatment for comparison to pre-treatment baseline recordings. Perceptual ratings and reliability measures were collected in the same manner as in 3.4.2 except raters adhered to a stricter "diagnostic" perceptual standard. Pre-to-post change in average of listener binary ratings were modeled to determine if change from pre-treatment to post-treatment was significantly different from zero for each participant, as in [25]. Multilevel models allow for the calculation of individual slopes without adjustment for multiple comparisons [26]. The models were fit in SAS PROC MIXED with restricted maximum likelihood estimation. Lastly, we quantified the effect magnitude (Busk and Serlin's $d_2$ [27]) for comparison to previous clinical trials for speech sound disorders, using $d_2 = 1$ as the customary threshold for clinical improvement [28].

## 4. Results

### 4.1. Hardware Validation: Audio and Rhotic Capture

We reviewed all 4,297 audio files recorded by ChainingAI to determine the number of utterances with complete audio capture of the target. 14% of utterances from one participant, 1130, experienced (detrimental) automatic signal processing to avoid clipping due to loud vocal volume that was not detectable in real time in the video conferencing audio signal. Separately, 4.3% of audio files were judged by ≥ one rater to not correspond to the complete written transcript, 1.0% of which were directly observed by the researcher to result from obvious user error with recording buttons, leaving 3.3% of utterances with insidious audio capture failure. Combined with the clipped files, ~7.4% of files were not fully captured by the system.

We reviewed 1,537 GMM-HMM boundaries (1107, 1111, 1112). Durations (Table 2) showed a large impact of GMM-HMM estimation for word-initial rhotics from treatment productions. GMM-HMM captured a median (q1, q3) of 7% (4%, 21%) of 1-ms samples of ground-truth rhotic interval word-initially, 82% (39%, 98%) word-medially, and 83% (60%, 100%) word-finally. Visual inspection indicated large differences word-initially were due to GMM-HMM estimating the offset boundary too early compared to the ground truth.

Table 2. *Duration (seconds) of rhotic-associated intervals by word context, reported as median (q1, q3).*

| Method | Initial | Medial | Final |
|---|---|---|---|
| GMM-HMM | .04 (.03, .10) | .30 (.13, .46) | .36 (.26, .49) |
| Ground Truth | .57 (.27, .79) | .33 (.17, .56) | .38 (.28, .55) |

### 4.2. Analytical Validation: PERCEPT/Rater Reliability

We reviewed reliability for 3,776 tokens from ChainingAI treatment sessions. Participant-level data appear in Figure 1 and Table 3. Gwet's chance-corrected agreement coefficient (γ) averaged for inter-human reliability at the level of the participant, was "moderate" ($\bar{\gamma}$ = .41, $\sigma_\gamma$ = .17); note the benchmarking of γ is relative to [29]. For comparison to previous literature [16], omnibus inter-rater ICC for our sample was 0.62 (95% CI [.60, .64]; two-way mixed effects, absolute agreement, multiple raters/ measurements). For pairwise F1 comparisons, the mean F1-score of human-human comparisons was 0.66 (0.12, [.38-.89]) versus 0.57 (0.16 [.327-.84]) for PERCEPT-human comparisons. As seen in Figure 1, PERCEPT-human performance was largely within the range of human-human performance for 4/5 of participants.

Table 3. *PERCEPT treatment performance (F1-score relative to clinician mode; [true derhotic, false rhotic | false derhotic, true rhotic], normalized by ground-truth) and inter-rater reliability of ground truth measures (γ, 95% CI, benchmark).*

| ID | F1-Score and Confusion Matrix | Interhuman Reliability (Gwet's γ) |
|---|---|---|
| 1107 | .55 [.29, .71 | .18, .82] | .23 [.18, .28], Fair |
| 1111 | .70 [.27, .73 | .08, .92] | .36 [.30, .43], Fair |
| 1112 | .62 [.46, .54 | .33, .67] | .50 [.45, .55], Moderate |
| 1121 | .81 [.54, .46 | .15, .84] | .65 [.69, .67], Good |
| 1130 | .39 [.18, .82 | .07, .93] | .31 [.26, .36], Fair |

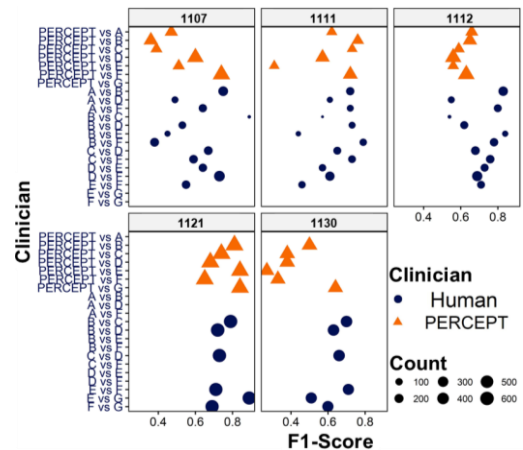

Figure 1. *Pairwise performance (F1-score) of PERCEPT is largely within the F1-score range of expert human raters.*

### 4.3. Clinical Validation: Response to Treatment

The means of the listener binary ratings (0 = derhotic, 1 = fully rhotic) for 4,315 rated untreated words elicited during the no-treatment baseline and post-treatment phases (Figure 2) were modeled. Average intra-rater reliability was "good" for these tokens ($\bar{\gamma}$ = .78, $\sigma_\gamma$ = .08, .66 ≤ $\bar{\gamma}$ ≤ .88) while omnibus inter-rater reliability was "moderate" ($\bar{\gamma}$ = .37, SE = .01, 95%CI[.36, .39]). Perceptual improvements were statistically different from zero for all 5 participants. 2 participants also demonstrated clinically significant standardized effect sizes ($d_2$ > 1): participant 1107 ($\hat{\gamma}$ = .47, SE = .021, $p$ <.0001, $d_2$ = 1.6) and 1121 ($\hat{\gamma}$ = 0.39, SE = .02, $p$ = <.0001, $d_2$ = 1.30). Three participants had effect sizes below the customary threshold of clinical significance: 1111 ($\hat{\gamma}$ = .18, SE = .020, $p$ <.0001, $d_2$ = .69), 1112 ($\hat{\gamma}$ = 0.13, SE = .03, $p$ = <.0001, $d_2$ = .36), and 1130 ($\hat{\gamma}$ = .13, SE = .02, $p$ = <.0001, $d_2$ = .40).

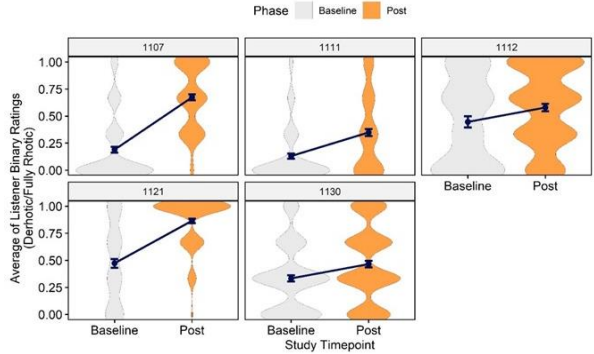

Figure 2. *Pre- to post-treatment change in untreated words show participants improved following the treatment. Bars represent 95% confidence interval of the mean.*

## 5. Discussion

To our knowledge this is the first application of the framework presented in [12, 13] for the prospective validation of clinical speech technology for speech sound disorders, including those of /ɹ/. It is our hope that this research will inform the long-term development of clinical speech technology such as ChainingAI.

### 5.1. Hardware Validation: Audio and Rhotic Capture

Our method successfully captured over 92% of recorded audio, and this implementation of a GMM-HMM forced alignment successfully captured, on average, 82%-83% of word medial and word final rhotics, but only 7% of word-initial rhotics. Participants were informally observed to be more successful when controlling the recording interface with a mouse versus a laptop track pad. In the future, we plan to explore key-press triggers for the recordings for users who may be interested in alternate input modalities, and to determine if participants can effectively monitor an input level visualization to avoid clipping. Future work can also reanalyze PERCEPT-R Classifier accuracy from the ground truth intervals to estimate the impact of GMM-HMM error on rhotic mispronunciation detection, and/or investigate whole-utterance analysis methods.

### 5.2. Analytical Validation: PERCEPT/Rater Reliability

The PERCEPT-R Classifier showed a participant-specific F1-score = .81 ($\sigma_x$= .10) in lab testing versus a strictly set ground truth [7]. However, Table 3 shows participant-specific F1-score = .61 ($\sigma_x$= .14). The high rate of false positives and low rate of false negatives at baseline in Table 3 may support the hypothesis that stimulable participants have more ambiguous /ɹ/ errors than participants in the PERCEPT-R Corpus. This is also supported by the observation that our rater reliability herein was lower than preliminary work using the same methods and raters.

The extent of variation seen in expert rater reliability (i.e., Gwet's, F1-score) was surprising. There are many possible explanations. First, many of these utterances were perceptually ambiguous as these participants, by definition, could say a fully rhotic /ɹ/ is some but not every utterance, due to difficulties with real-time coarticulation or underarticulation rather than a wholly derhotic motor plan. There seems to be a subset of clinical speech associated with ambiguous perception for expert listeners, at least in /ɹ/ speech sound disorders (see also: [15, 16]), but likely in other speech disorders as well. It may be that a gold-standard perceptual ground truth is more difficult to assign in clinical speech than typical speech, and the ability of a singular metric based on that ground truth (i.e., F1-score) to serve as a universal proxy for clinical potential is called into question by the results shown in Figure 3. Future work by multiple research teams should evaluate this possibility further.

Another source of variation between human listeners is that, for in-treatment tokens, clinicians likely differ in their clinical style regarding motivation of participants, particularly through the acceptability of marginal productions. Lastly, clinicians heard these utterances in random order versus the order that would naturally occur in a treatment session. It is a common clinical strategy to accept more marginal productions early in treatment and then adopt a stricter perceptual threshold as the learner's productions improve, but use of this strategy among raters may have differed in this task. Relatedly, the PERCEPT-R Classifier was operationalized not to adapt in nature within/between sessions, which might explain differences between baseline F1 and treatment F1, and could be a goal of future research.

### 5.3. Clinical Validation: Response to Treatment

Even with the wide range of F1-scores for the present study—including F1-scores at or below chance (Table 3)—all participants demonstrated statistically significant improvement in untreated words after ten 40-minute sessions assisted with ChainingAI. This finding raises questions about what aspects of clinical feedback interact with learner characteristics, such as self-monitoring skill, focus, and motivation, to impact long-term treatment gains. Further work with more participants can investigate paradoxes that may be at hand: that the individuals most suited for independent practice (i.e., those with occasional/marginal correct /ɹ/) may be most prone to low rating reliability, but perhaps can improve with feedback that does not always match the judgment of any one clinician.

### 5.4. Limitations

There are limitations in both the procedure and reporting of this study. For one, we did not use cross validated reporting for personalization for the first three participants. Secondly, this work does not isolate the effect of ChainingAI or report customary single case experimental analyses, which will be reported in subsequent work due to space limitations.

## 6. Conclusions

We demonstrate prospective validation of the PERCEPT-R Classifier: hardware validation (92.6% audio capture), analytical validation (F1-score $\bar{x}$ = .61, $\sigma_x$= .14, n = 5), and clinical validation (statistically-significant pre- to post-treatment change in untreated words following 10 sessions of combined human-ChainingAI treatment for all participants, with clinically significant changes for two of five participants). Participant improvement despite low classifier—and human—agreement may indicate that metrics based on potentially ambiguous perceptual ground truth cannot singularly index clinical potential of speech technologies for specific learners.

## 7. Acknowledgements

This research was supported through an internal grant (CUSE II-14-2021; J. Preston, PI) and computational resources (NSF ACI-1341006; NSF ACI-1541396) from Syracuse University.